\documentstyle[aps,prl,epsf,floats]{revtex}
\def \Pt {{P}_{T}}
\def \Et {{E}_{T}}
\def \met  {\,/\!\!\!\!E_{T}}
\def \mtop {M_{top}}
\def \Ht {H_{T}}
\def\r#1{\ignorespaces $^{#1}$}
\newcommand{\ttbar}{t\bar{t}}

\begin{document}                                             

%\begin{flushright}
%CDF/DOC/TOP/CDFR/4008 \\
%FERMILAB-PUB-97/122-E \\
%Accepted for publication by PRL 26 Jan 1997
%\end{flushright}

\title{
{\bf Measurement of the Top Quark Mass and $\ttbar$ Production Cross Section
from Dilepton Events at the Collider Detector at Fermilab}
}

%\renewcommand{\baselinestretch}{2}

%----------------
%  author list
%----------------
%\input{cdf_runi_authors.tex}
\author{
\font\eightit=cmti8
\hfilneg
\begin{sloppypar}
\noindent
F.~Abe,\r {17} H.~Akimoto,\r {39}
A.~Akopian,\r {31} M.~G.~Albrow,\r 7 A.~Amadon,\r 5 S.~R.~Amendolia,\r {27} 
D.~Amidei,\r {20} J.~Antos,\r {33} S.~Aota,\r {37}
G.~Apollinari,\r {31}
% ...
\end{sloppypar}
\vskip .026in
\begin{center}
(CDF Collaboration)
\end{center}
\vskip .026in
\begin{center}
\r 1  {\eightit Argonne National Laboratory, Argonne, Illinois 60439} \\
\r 2  {\eightit Istituto Nazionale di Fisica Nucleare, University of Bologna,
I-40127 Bologna, Italy} \\
\r 3  {\eightit Brandeis University, Waltham, Massachusetts 02254} \\
\r 4  {\eightit University of California at Los Angeles, Los 
Angeles, California  90024} \\  
\r 5  {\eightit University of Chicago, Chicago, Illinois 60637} \\
\r 6  {\eightit Duke University, Durham, North Carolina  27708} \\
\r 7  {\eightit Fermi National Accelerator Laboratory, Batavia, Illinois 
60510} \\
\r 8  {\eightit University of Florida, Gainesville, FL  32611} \\
\r 9  {\eightit Laboratori Nazionali di Frascati, Istituto Nazionale di Fisica
               Nucleare, I-00044 Frascati, Italy} \\
\r {10} {\eightit University of Geneva, CH-1211 Geneva 4, Switzerland} \\
\r {11} {\eightit Harvard University, Cambridge, Massachusetts 02138} \\
\r {12} {\eightit Hiroshima University, Higashi-Hiroshima 724, Japan} \\
\r {13} {\eightit University of Illinois, Urbana, Illinois 61801} \\
\r {14} {\eightit Institute of Particle Physics, McGill University, Montreal 
H3A 2T8, and University of Toronto,\\ Toronto M5S 1A7, Canada} \\
\r {15} {\eightit The Johns Hopkins University, Baltimore, Maryland 21218} \\
\r {16} {\eightit Institut f\"{u}r Experimentelle Kernphysik, 
Universit\"{a}t Karlsruhe, 76128 Karlsruhe, Germany} \\
\r {17} {\eightit National Laboratory for High Energy Physics (KEK), Tsukuba, 
Ibaraki 305, Japan} \\
\r {18} {\eightit Ernest Orlando Lawrence Berkeley National Laboratory, 
Berkeley, California 94720} \\
\r {19} {\eightit Massachusetts Institute of Technology, Cambridge,
Massachusetts  02139} \\   
\r {20} {\eightit University of Michigan, Ann Arbor, Michigan 48109} \\
\r {21} {\eightit Michigan State University, East Lansing, Michigan  48824} \\
\r {22} {\eightit University of New Mexico, Albuquerque, New Mexico 87131} \\
\r {23} {\eightit The Ohio State University, Columbus, OH 43210} \\
\r {24} {\eightit Osaka City University, Osaka 588, Japan} \\
\r {25} {\eightit Universita di Padova, Istituto Nazionale di Fisica 
          Nucleare, Sezione di Padova, I-35131 Padova, Italy} \\
\r {26} {\eightit University of Pennsylvania, Philadelphia, 
        Pennsylvania 19104} \\   
\r {27} {\eightit Istituto Nazionale di Fisica Nucleare, University and Scuola
               Normale Superiore of Pisa, I-56100 Pisa, Italy} \\
\r {28} {\eightit University of Pittsburgh, Pittsburgh, Pennsylvania 15260} \\
\r {29} {\eightit Purdue University, West Lafayette, Indiana 47907} \\
\r {30} {\eightit University of Rochester, Rochester, New York 14627} \\
\r {31} {\eightit Rockefeller University, New York, New York 10021} \\
\r {32} {\eightit Rutgers University, Piscataway, New Jersey 08855} \\
\r {33} {\eightit Academia Sinica, Taipei, Taiwan 11530, Republic of China} \\
\r {34} {\eightit Texas A\&M University, College Station, Texas 77843} \\
\r {35} {\eightit Texas Tech University, Lubbock, Texas 79409} \\
\r {36} {\eightit Istituto Nazionale di Fisica Nucleare, University of Trieste/
Udine, Italy} \\
\r {37} {\eightit University of Tsukuba, Tsukuba, Ibaraki 315, Japan} \\
\r {38} {\eightit Tufts University, Medford, Massachusetts 02155} \\
\r {39} {\eightit Waseda University, Tokyo 169, Japan} \\
\r {40} {\eightit University of Wisconsin, Madison, Wisconsin 53706} \\
\r {41} {\eightit Yale University, New Haven, Connecticut 06520} \\
\end{center}
}
                      
\maketitle

\begin{abstract}
%---------------
We present an analysis of dilepton events originating from  ${t\bar{t}}$ 
production in ${\bar{p}p}$ collisions  at $\sqrt{s}=1.8$ TeV at the Fermilab 
Tevatron Collider. The sample corresponds to an integrated luminosity of 
$109\pm 7 ~{\rm pb^{-1}}$. We observe 9 candidate events, with an estimated 
background of $2.4 \pm 0.5$ events. We determine the mass of the top quark 
to be
$M_{top} = 161\pm 17 ({\rm stat.})\pm 10 ({\rm syst.}) ~{\rm GeV}/c^{2}$.
In addition we measure a ${t\bar{t}}$ production cross section of
8.2$^{+4.4}_{-3.4}$~pb (where $M_{top} = 175 ~{\rm GeV}/c^{2}$ has been 
assumed for the acceptance estimate).


\end{abstract}                           

%=====
\narrowtext
\twocolumn
%=====         

We report here on a measurement of the $t\bar{t}$ production 
cross section and top quark mass in the dilepton channel
with the Collider Detector at Fermilab (CDF). The data sample 
corresponds to a total integrated luminosity of $109\pm 7 ~{\rm pb^{-1}}$.
This analysis considers dilepton events originating predominantly from 
$t\bar{t} \rightarrow W^+bW^-\bar{b}
\rightarrow (\ell^+\nu b)(\ell^-\bar{\nu}\bar{b})$, with $\ell = e $ or $\mu$.
A subset of these events corresponding to a data sample 
of 67 pb$^{-1}$  supported the discovery of the top quark
\cite{disccdf,discd0}.

Recently both the CDF and D0 collaborations have
updated their measurements of the $t\bar{t}$ production 
cross section\cite{xscdf,xsd0} and top quark mass\cite{mtcdf,mtd0}
using the ``lepton plus jets'' channel:                 
$t\bar{t} \rightarrow W^+bW^-\bar{b}
\rightarrow (\ell^+\nu b)(q\bar{q}'\bar{b})$, with $\ell = e $ or $\mu$,
which has larger statistics.
The consistency of the measurements presented here with those in the lepton 
plus jets channel is an important confirmation 
that these two
orthogonal sets of events both originate from the same heavy top production
process.

A description of the CDF detector can be found 
in Ref.\cite{detector}. The coordinate system and various quantities used
throughout this paper are defined in \cite{eta}.
The momenta of the charged leptons are measured with the 
central tracking chamber in a 1.4-T superconducting  solenoidal magnet.  
Electromagnetic and hadronic calorimeters surrounding the tracking chambers 
are used to identify and measure the energies of electrons and jets.  
Muons are identified with drift chambers located outside the calorimeters.  
A three-level trigger selects high transverse momentum ($\Pt$) 
electrons and muons.

% event selction
%----------------
The event selection is very similar to the previous analysis described in 
Ref.\cite{disccdf,evidence}. We require two high-$\Pt$ 
($\Pt > 20~{\rm GeV}/c$), oppositely charged leptons ($e$ or $\mu$)
in the central pseudorapidity region ($|\eta|< 1.0$) with at least one of them 
well isolated from nearby tracks and calorimeter activity. 
We reject $Z\rightarrow \ell^{+}\ell^{-}X$ events by requiring the dilepton 
invariant mass, $M_{ee}$ or $M_{\mu\mu}$, to be outside the interval 
75-105 GeV/$c^2$.
If there is a high transverse energy ($\Et> 10$~GeV) isolated photon present, 
the event is removed if consistent with a radiative $Z$ decay.
We also require that there be at least two jets with measured 
$\Et > 10$~GeV in the pseudorapidity range  $|\eta|<2.0$, as expected from 
the presence of two $b$ quarks in a $t\bar{t}$ event. 
The signature of the two neutrinos in the decay is missing energy transverse 
to the beam direction ($\met$); we require $|\met|>25$~GeV. 
The $\met$ is corrected for non-uniformities in calorimeter response and 
absolute energy scale, and  for high-$\Pt$ muons.
To ensure that the $\met$ is not due to mismeasurements of the energies of 
the leptons or jets, we require $|\met| >$ 50~GeV if the azimuthal angle 
between the direction of the $\met$ vector and the nearest
lepton or jet($j$), $\Delta\phi(\met,\ell~{\rm or}~j)$, is less than 
20$^{\circ}$\cite{evidence}.

\begin{figure}                                      
\epsffile[36 100 540 684] {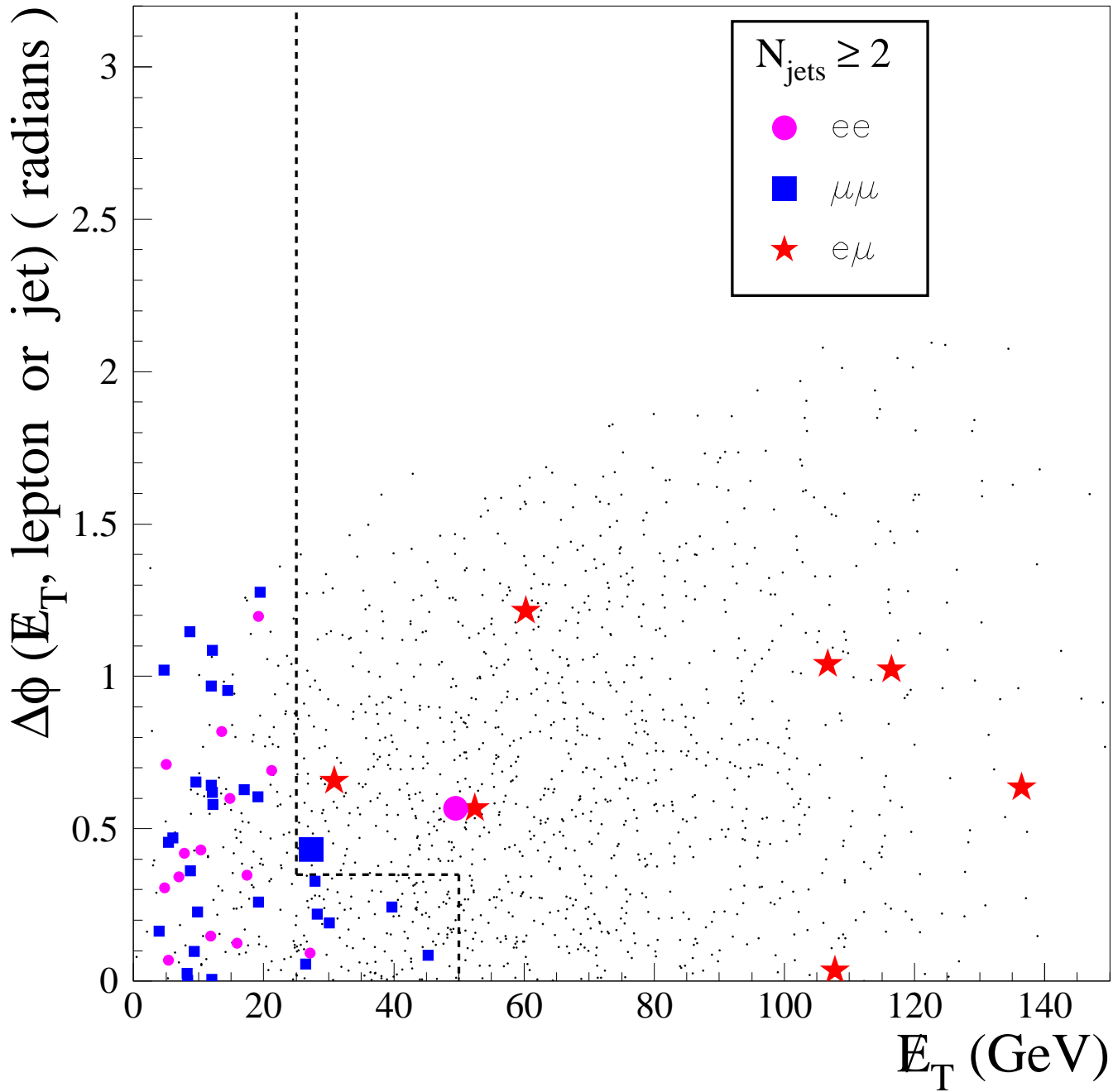}
\caption{Azimuthal angle between $\met$ and the nearest lepton or jet, 
  versus $|\met|$ for events with two leptons and two jets.
  The dashed line represents the $\met$ cut.
  The small dots are for $t\bar{t}$ Monte Carlo for
  $M_{top}$ = 175 GeV/$c^2$ and correspond to an integrated luminosity of
  about $24 ~{\rm fb}^{-1}$. The larger symbols represent the data.}
\label{fig:dil_data}
\end{figure}

The distribution of $\Delta\phi(\met,\ell~{\rm or}~j)$ versus $|\met|$
is shown in Figure~\ref{fig:dil_data} for dilepton events that pass
the invariant mass and two-jet requirements. Superimposed is
the expected distribution from  the HERWIG\cite{Herwig} 
$t\bar{t}$ Monte Carlo program for $M_{top}$ = 175 GeV/$c^2$
followed by a detector simulation\footnote{The top mass value of 
175 GeV/$c^2$ is used to be consistent with CDF's most precise mass
measurement, that from the lepton plus jets channel\cite{hadronic}.}.
We find nine candidate events in the signal region: 
seven $e\mu$, one  $\mu\mu$ and one $ee$ event.  

%efficiency and acceptance
%-------------------------
The identification efficiencies for single leptons are measured from
$Z\rightarrow \ell^{+}\ell^{-}$ events in the data and are found to be 
$91\%$ for muons and $83\%$ for electrons\cite{mark}.
The acceptance for $t\bar t$ decays to pass all the selection criteria,
including lepton identification, is the average of the results
from the HERWIG and PYTHIA\cite{Pythia} Monte Carlo
programs followed by a detector simulation.
We find for $M_{top}$ = 175 GeV/$c^2$ that $(0.74 \pm 0.08)\%$ of all 
$t\bar t$ decays pass the above dilepton selection criteria.
The uncertainty is dominated by the 
differences between the event generators and by the systematic
uncertainties in the detector simulation.
The acceptance increases by 35\% as $M_{top}$  
increases from 150 to 200 GeV/$c^2$.
In $(86\pm 2)\%$ of the dilepton events passing all selection criteria
both leptons come directly from the decays of the $W$ bosons.      
The remainder consists mostly of events in which one of the $W$ bosons decays
to a $\tau$ lepton, which in turn decays to an electron or muon.
Of the dilepton events, we expect $(58\pm 2)\%$ to be $e\mu$, 
$(27\pm 1)\%$ $\mu\mu$ and $(15\pm 1)\%$ $ee$, where the uncertainty
is statistical only.                                           
Using a  $\ttbar$ production cross section value of 5.5 pb,
consistent with recent theoretical calculations\cite{theory} that
assume $M_{top}$=175 GeV/$c^2$,  we expect to observe 4.4 signal events.

% backgrounds
%------------   
As backgrounds we consider standard model processes, other than $\ttbar$,
that can result in dilepton final states. The main sources are
Drell-Yan ($Z^*/\gamma \to ee,\mu\mu$), $Z \rightarrow \tau\tau$ and $WW$ 
production.
If these events contain additional jets from QCD radiation plus  $\met$, either
from real neutrinos or from energy mismeasurement, they may satisfy our 
selection criteria. 
Background contributions from radiative $Z$ bosons and from $b\bar b$, 
$WZ$, $ZZ$ and $Wb\bar b$ production are estimated to be small.
Additional sources are processes with a real lepton and a jet or a track
faking a second lepton, and processes in which
mismeasured muon tracks can result in an overestimate of the 
$\met$ in the event. This latter background is not relevant for electrons
because the electron energy is measured in the calorimeters.
We estimate the background from Drell-Yan production, fake leptons
and mismeasured tracks from the data;
the other backgrounds are calculated using Monte Carlo simulations.
The background contributions from the different sources are listed in
Table I.~\ref{table:tbck} The errors include both systematic and
statistical uncertainties. Of the $2.4\pm 0.5$ total background events
estimated, $0.8\pm 0.2$ are expected in the $e\mu$ channel, which does not have
a contribution from Drell-Yan production, 
the dominant background source in the $ee$ and
$\mu\mu$ channels.

%%%%%relaxing the cuts:                        

When relaxing the two-jet requirement in the data selection,
we find eight dilepton events with zero
jets and eleven with one jet, while we expect $8\pm 2$ and $7\pm 2$ events,
respectively, from both background and $\ttbar$. 
The $\ttbar$ contribution is small; using the measured 
$\ttbar$ production cross section in this channel (see below) we expect
$0.03\pm 0.02$ dilepton events in the zero jet sample and $1.1\pm 0.5$
events with one jet. Of the eleven events in the data with one jet, 
four are dimuons, three of which show the expected characteristics of the
mismeasured muon track background. The expectation from this background
is $1.4\pm 1.5$ events. Five of the remaining seven events are $e\mu$.
One of these five events has a jet tagged as a $b$ quark (see below). 
The expected number of $t\bar t$ events in the one jet sample with the jet
tagged as a $b$ quark is about 0.2.

%\begin{table}[b]
\begin{table}
\begin{center}
\caption{Expected dilepton backgrounds in $109\pm 7 ~{\rm pb^{-1}}$.}
\begin{tabular}{cc} %\hline %\hline
Background type & Expected \# of events \\ \hline
 Drell-Yan      &   $0.61 \pm 0.30 $ \\ 
 $Z\to\tau\tau$ &   $0.59 \pm 0.14 $ \\ 
 Fake leptons   &   $0.37 \pm 0.23 $ \\ 
 $WW$           &   $0.36 \pm 0.11$ \\ 
 Mismeasured muon tracks & $0.3\pm 0.3$ \\ 
 $b\bar{b}$     &   $0.05 \pm 0.03 $ \\ 
Other(radiative $Z,Wb\bar{b},WZ,ZZ$) & $0.1\pm 0.1$ \\ \hline
 {\bf Total}    &   $2.4 \pm 0.5$ \\ % \hline %\hline 
\end{tabular}
\end{center}
\label{table:tbck}
\end{table}

We find two events in the data that satisfy all selection 
criteria, except for the opposite sign requirement on the lepton charge.
We expect $0.37\pm 0.23$ same sign dilepton plus two-jet events from fake 
lepton background
and $0.24\pm 0.11$ from $t\bar{t}$. Without any jet requirements 
these estimates are $2.6\pm 1.9$ and $0.29\pm 0.13$, respectively.
There are no events in the data with two same sign leptons
and 0 or 1 jets that satisfy all the other selection criteria.

Events originating from $t\bar t$ decays are characterized by the presence 
of jets originating from $b$ quarks. Four of the nine candidate events 
have one jet tagged as a $b$ quark, as evidenced by the presence of a 
secondary vertex in the silicon vertex detector (SVX tag)~\cite{evidence}.
Two of these four jets are also tagged by a soft lepton from the semileptonic 
decay of the $b$ quark (SLT tag) ~\cite{evidence}. No jets are found to be 
tagged by the SLT method alone.
In a sample of seven $t\bar{t}$ events, $4.3\pm 0.4$ jets are expected to be 
tagged by at least one of the two tagging methods.  If all nine candidates 
were from background processes, the number of expected $b$-tagged jets would 
be $0.7\pm 0.2$.

Using the nine observed dilepton events, $2.4\pm 0.5$ of which are estimated 
to be background, an overall acceptance of $(0.74\pm 0.08)\%$           
(for $M_{top}$ = 175 GeV/$c^2$), and an integrated luminosity of 
$109\pm 7 {\rm pb}^{-1}$, we calculate the  $t\bar{t}$ production 
cross section to be 8.2$^{+4.4}_{-3.4}$~pb.

%---------------------------------------------------------------- 
% mass
%-------
With $t\bar t$ dilepton decays, reconstructing the top quark mass 
from the measured final state is a kinematically under-constrained problem 
due to the presence of two neutrinos.   

\begin{figure}
%\epsfxsize=8.0cm                  
%\epsfysize=4.5cm                  
%\epsffile[36 162 540 684] {kin_v4.ps}
\epsffile[36 162 540 684] {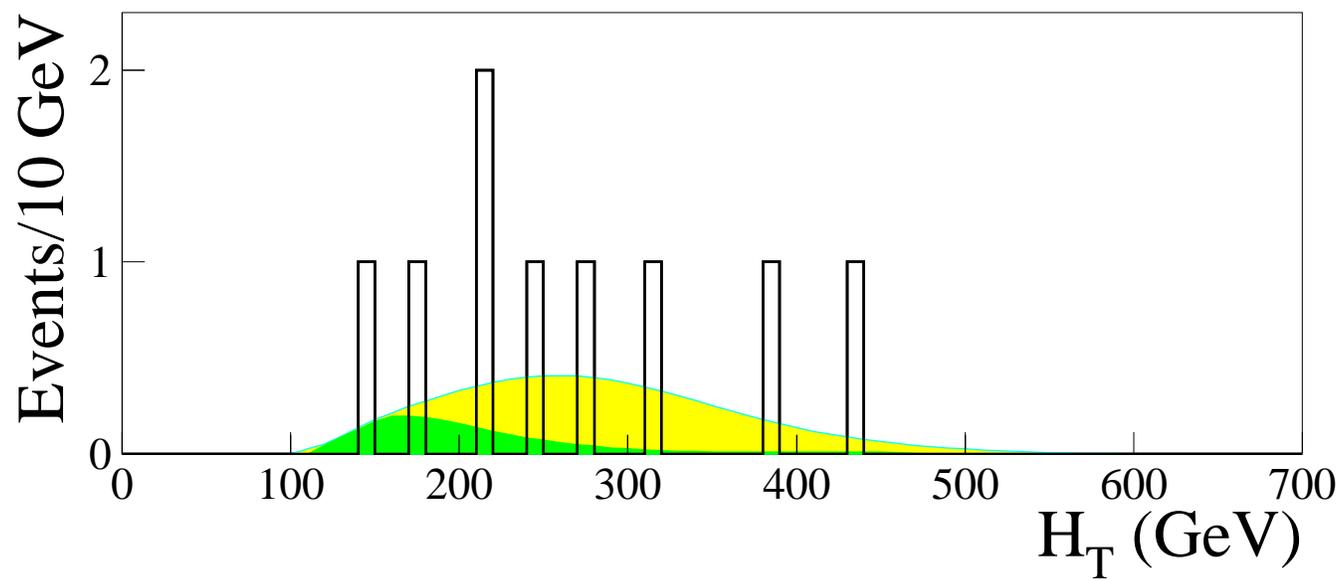}
\caption{Comparison of $\Ht$ for the candidate events (histogram) and
  the expectation  from $t\bar t$ production ($\mtop$ = 175 GeV/$c^2$) 
  plus background (lighter shaded area). The background distribution alone
  is represented by the darker shaded area.} 
\label{fig:kin}
\end{figure}

To increase the purity of the sample for the mass analyses, we require
$\Ht> 170$~GeV.  $\Ht$ is defined as the sum of the $\Pt$'s of
the two leptons ($\Et$ for electrons), 
the $\Et$'s of the two highest $\Et$ jets and the $|\met|$. 
For the nine candidate events the $\Ht$ distribution
is shown in Figure~\ref{fig:kin}, together with
the expectations from a Monte Carlo calculation
for $\mtop$ = 175 GeV/$c^2$ and background. 
The efficiency of the $\Ht> 170$~GeV
cut is about $95\%$ for $M_{top}$ = 175 GeV/$c^2$.
After this cut, eight of the nine dilepton candidate events survive
and the background is reduced to $1.3\pm 0.3$ events.

We present  two methods to determine the top quark mass.
The first method takes advantage of the correlation between the energy of 
the jets from $b$ quarks in top quark decays and the top quark mass.
From Monte Carlo simulation of $t\bar t$ decays
we find that the mean energy of the two highest $\Et$ jets increases 
linearly with $M_{top}$ with a slope of 0.5.
We obtain the most probable value of $M_{top}$ for our set of events by 
comparing the observed distribution of  jet energies with Monte
Carlo distributions (templates)  for different top quark masses.  
We use the two highest $\Et$ jets in the event.
The jet energies are corrected for non-uniformities in calorimeter response 
and absolute energy scale. The  templates for $t\bar t$ decays are obtained
from large samples of  Monte Carlo events generated with HERWIG, 
and reconstructed with the same analysis programs as the data.   
Templates are also obtained for the background processes discussed above.
The top quark mass is estimated by performing a maximum likelihood fit
of the jet energy distribution from the data to 
a combination of $t\bar t$ and background templates.
The amount of background is constrained to the expected value within 
its uncertainty. From the resulting likelihood values, ${\cal{L}}_m$, for 
each assumed top quark mass $m$, the negative logarithms, $-\ln({\cal{L}}_m)$,
are fit with a third order polynomial and the value of the mass 
corresponding to the  minimum of the polynomial is obtained.    
The statistical uncertainties are determined by the values of the mass that 
give an increase of 0.5 in $-\ln({\cal{L}}_m)$ from the minimum.

We have tested the fitting method with many Monte Carlo samples,
with $\mtop$ in the range between 100 and 240 GeV/$c^2$, and with the same
statistics as the data.
Included in these samples is the expected number of background events.
The mean of the distribution of the estimated top quark mass for each sample
agrees with the generated mass within 2~GeV/$c^2$.  
The average statistical uncertainties derived from the likelihood fits are  
consistent with the spread of the distributions, about 21~${\rm GeV}/c^2$, and
are not strongly dependent on the top quark mass.   

\begin{figure}
\epsffile[36 162 540 684] {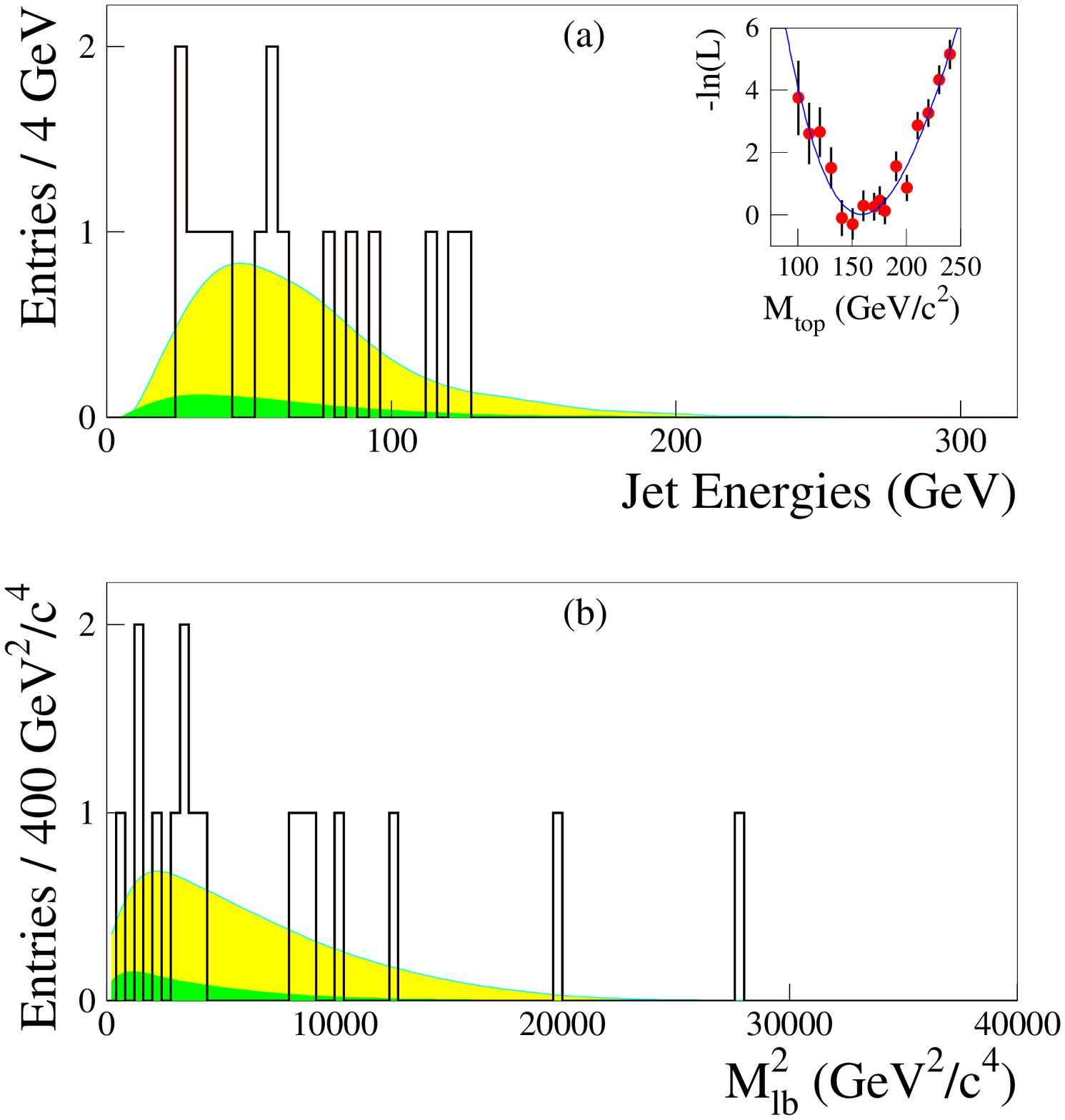}
\vspace*{2cm}
\caption{(a) Jet energy distribution of the two highest $\Et$ jets 
  for the dilepton events (histogram).  Superimposed is the same distribution
  for $t\bar t$ Monte Carlo ($\mtop$=160 GeV/$c^{2}$) plus background (lighter
  shade), and the background alone (darker shade). 
  The $\ttbar$ plus background distribution is normalized to the data.
  In the inset we show the $-\ln({\cal{L}}_m)$ fit as a function of $\mtop$
  (the minimum has been offset to be at zero).
  (b) Distribution of ${M_{\ell b}^2}_{min}$ (histogram). Superimposed is the 
  expectation from $t\bar t$ Monte Carlo ($\mtop$=160 GeV/$c^{2}$) plus 
  background (lighter shade), and background alone (darker shade).}
\label{fig:data}
\end{figure}

Figure~\ref{fig:data}(a) shows the jet energy distribution of the two 
highest $\Et$ jets for the eight events in the data overlayed
with a template obtained from a combination of $t\bar t$ Monte Carlo
($\mtop$ = 160 GeV/$c^2$) and background in the ratio of 6.7 to 1.3. 
The distribution of the background alone is also shown. 
The inset shows the polynomial fit to the $-\ln({\cal{L}}_m)$ values
versus top quark mass. With this method we measure 
$M_{top} = 159\pm 23({\rm stat.}) \pm 11({\rm syst.}) {\rm GeV}/c^2$.

%%%%%%% MLB STUFF %%%%%%%

The second method uses the invariant mass $M_{\ell b}$ of
the charged lepton $\ell$ and the $b$ quark.  In the decay
of top quarks ($t\to Wb$), the energy of the $b$ quark
in the rest frame of the $W$ boson, $E_b$, is a constant
($M_{top}^2 = M_W^2 + M_b^2 + 2 M_W E_b$).
With the subsequent semileptonic decay $W\to\ell\nu$,
$E_b$ can be obtained from the invariant mass $M_{\ell b}$
of the charged lepton $\ell$ and the $b$ quark, and their
opening angle $\cos\theta_{\ell b}$.
For a sample of $\ttbar$ dilepton events, $M_{top}$ is related to
$M_{\ell b}$ as
$M_{top}^2 = M_W^2 + {2\langle M_{\ell b}^2\rangle \over 
	      		1-\langle \cos\theta_{\ell b}\rangle}$,
where terms that include lepton and $b$ quark masses
have been neglected.
%with a combined error of less than 0.2\% on $M_{top}$.
$\langle M_{\ell b}^2\rangle$ and $\langle \cos\theta_{\ell b}\rangle$
are the mean values of $M_{\ell b}^2$ and $\cos\theta_{\ell b}$ in
the sample.
In the standard model tree-level calculation,
$\langle \cos\theta_{\ell b}\rangle = M_W^2/(M_{top}^2 + 2 M_W^2)$,
resulting in
$M_{top}^2 = \langle M_{\ell b}^2\rangle +
	\sqrt{M_W^4 + 4 M_W^2 \langle M_{\ell b}^2\rangle
		    + \langle M_{\ell b}^2\rangle^2}$.

In each dilepton $t\bar t$ candidate, one of the two lepton-jet
combinations $(\ell^+j_1,\ell^-j_2)$, $(\ell^+j_2,\ell^-j_1)$
corresponds to the correct $(\ell^+b,\ell^-\bar b)$ assignment.
We select the combination with the smallest sum of invariant masses,
resulting in two values of ${M_{\ell b}^2}_{min}$ per event.      
The probability of picking up the correct combination is
in the range of 55\% to 75\%, depending on $M_{top}$.
The distribution of ${M_{\ell b}^2}_{min}$ for the eight candidate
events is shown in Figure~\ref{fig:data}(b) 
together with the expectation for $M_{top}$ = 160 GeV/$c^2$.
The mean value of ${M_{\ell b}^2}_{min}, ~(7.2\pm 1.9)\times 10^3$ ~GeV/$c^2$,
is used with                         
the linear mapping function 
$\langle M_{\ell b}^2\rangle = C_0 + C_1 \langle {M_{\ell b}^2}_{min}\rangle$
(with $C_0 = (-2.90\pm 0.26)\times 10^{3},~C_1 = 1.57\pm 0.03$)
to obtain the value of $\langle M_{\ell b}^2\rangle$ corresponding to
standard model $t\bar t$ production and decay.
The mapping function is determined from $t\bar t$ Monte Carlo events 
generated with HERWIG and a CDF detector simulation.
It accounts for selection biases, incorrect lepton-jet  
combinations, and jet energy mismeasurements.
We obtain $M_{top} = 163\pm20({\rm stat.})\pm9({\rm syst.})$ GeV/$c^2$.

We combine the two mass results taking into account the correlations in the
uncertainties to obtain a single value for the top quark mass,
$M_{top} = 161\pm 17 ({\rm stat.})\pm 10 ({\rm syst.}) ~{\rm GeV}/c^{2}$.
The major contributions to the systematic uncertainty
are due to the uncertainty in the jet energy scale and in the shape
of the background distributions.                         

%conclusion
%----------
In conclusion, in the dilepton channel we find nine candidate events 
consistent with originating from $\ttbar$ production. The estimated background
is $2.4\pm 0.5$ events. We measure the $\ttbar$ production cross section to 
be $8.2^{+4.4}_{-3.4}$~pb for $M_{top} = 175 ~{\rm GeV}/c^{2}$. The measured 
top quark mass is 
$M_{top} = 161\pm 17 ({\rm stat.})\pm 10 ({\rm syst.}) ~{\rm GeV}/c^{2}$,
consistent with the CDF measurement in the lepton plus jets channel of
$175.9\pm 4.8 ({\rm stat.})\pm 4.9 ({\rm syst.}) ~{\rm GeV}/c^{2}$
\cite{hadronic}.

We thank the Fermilab staff and the technical staffs of the
participating institutions for their contributions.  This work was
supported by the U.S. Department of Energy and National Science Foundation;
the Italian Istituto Nazionale di Fisica Nucleare; the Ministry of Science,
Culture, and Education of Japan; the Natural Sciences and Engineering Research
Council of Canada; the National Science Council of the Republic of China;
and the A. P. Sloan Foundation.

%-----------------------
%  Bibliography
%-----------------------

\onecolumn

%------------------------
%  FIGURES (3 in total)
%------------------------

\newpage
%\begin{center}
%\end{center}

\newpage                                                  

\newpage


\begin{thebibliography}{99}
              
\bibitem{disccdf}  
  F. Abe $\it et~ al.,$ Phys. Rev. Lett. {\bf 74}, 2626 (1995).

\bibitem{discd0}  
  S. Abachi $\it et~ al.,$ Phys. Rev. Lett. {\bf 74}, 2632 (1995).
                   
\bibitem{xscdf}                                                   
  F. Abe $\it et~ al.,$ Phys. Rev. Lett., to be published.
                   
\bibitem{xsd0}  
  S. Abachi $\it et~ al.,$ Phys. Rev. Lett., {\bf 79}, 1203 (1997).
                   
\bibitem{mtcdf}                                                   
  F. Abe $\it et~ al.,$ Phys. Rev. Lett., to be published.
                   
\bibitem{mtd0}  
  S. Abachi $\it et~ al.,$ Phys. Rev. Lett., {\bf 79}, 1197 (1997).
                   

\bibitem{detector} 
  F. Abe $\it et~ al.,$  Nucl. Instr. Meth. Phys. Res. A {\bf271}, 387 (1988).

\bibitem{eta} 
  In the CDF coordinate system, $\theta$ and $\phi$ are the polar
  and azimuthal angles, respectively, with respect to the proton
  beam direction ($z$ axis). The pseudorapidity $\eta$ is defined as
  $-\ln \tan(\theta/2)$.
  The transverse momentum of a particle is $\Pt = P \sin\theta$.
  The analogous quantity using calorimeter energies, defined as
  $\Et = E \sin\theta$, is called transverse energy.
  The missing transverse energy, $\met$, is defined as
  $-\sum \Et^i \,\hat{n}_i$, where $\hat{n}_i$ is the unit vector in the 
  transverse plane pointing from the interaction point to
  the energy deposition in calorimeter cell $i$.

\bibitem{evidence} 
  F. Abe $\it et~ al.,$  Phys. Rev. D {\bf 50}, 2966 (1994); 
  Phys. Rev. Lett. {\bf 73}, 225 (1994). 

\bibitem{Herwig} 
  G. Marchesini and B.~R. Webber, Nucl. Phys. {\bf 310}, 461 (1988); 
  G. Marchesini $\it et~ al.$, Comput. Phys. Commun. {\bf 67}, 465~(1992).
  We use HERWIG V5.6.
                                                 
\bibitem{hadronic} 
  F. Abe $\it et~ al.,$ Report No. Fermilab-Pub-97/284-E, to be published.
                           
\bibitem{mark} 
  M. C. Kruse, Ph.D. Thesis, Purdue University, USA (1996).

\bibitem{Pythia} 
  T.~Sj\"{o}strand, Comput. Phys. Commun. {\bf 82}, 74 (1994). 
  We use PYTHIA V5.7.

\bibitem{theory} 
  E. Berger and H. Contopanagos, Phys. Rev. D {\bf 54}, 3085 (1996);
  S. Catani, M. Mangano, P.Nason, and L. Trentadue,
  Phys. Lett. B {\bf 378}, 329 (1996);
  E. Laenen, J. Smith and W. L. van Neerven, Phys. Lett. B {\bf 321}, 
  254 (1994). 

\end{thebibliography}
\end{document}